\documentclass[11pt]{amsart}

\usepackage[hmargin={3cm,3cm},vmargin={3cm,3cm},includehead]{geometry}

\usepackage{cite}
\usepackage{amsmath,amssymb}
\usepackage{amsfonts}
\usepackage{mathrsfs}
\usepackage{epic}
\usepackage{eepic}

\usepackage{graphicx}

\newcommand{\ds}{\displaystyle}
\def\EXP{\textrm{{\large e}}}
\def\eqref#1{(\ref{#1})}
\newcommand{\nop}{\boldsymbol{n}}
\newcommand{\e}{\boldsymbol{e}}

\newcommand{\bos}{\boldsymbol{a}}
\newcommand{\kop}{\boldsymbol{k}}

\newcommand{\ii}{\mathsf{i}}
\newcommand{\bb}{b}

\newcommand{\Rop}{\boldsymbol{R}}
\newcommand{\rop}{\boldsymbol{r}}

\newcommand{\alg}{\mathcal{A}}

\newcommand{\Bigpsi}[3]{\phantom{\Psi}_2 \kern -.05em
\Psi_2\left(\genfrac{}{}{0pt}{}{#1}{#2}\biggl|#3\right)}

\font\cyr=wncyr10 % cyrillic font for Milnor's Lobachevski function.
\newcommand{\Lb}{\operatorname{\mbox{\cyr L}}}

\begin{document}

\title[]{Classical integrable field theories in discrete $2+1$ dimensional space-time.}
\author{Sergey M. Sergeev}
\address{Faculty of Informational Sciences and Engineering,
University of Canberra, Bruce ACT 2601}
\email{Sergey.Sergeev@canberra.edu.au}

\begin{abstract}
We study ``circular net'' (discrete orthogonal net) equations for
angular data generalized by external spectral parameters. A
criterion defining physical regimes of these Hamiltonian equations
is the reality of Lagrangian density. There are four distinct
regimes for fields and spectral parameters classified by four
types of spherical or hyperbolic triangles. Non-zero external
spectral parameters provide the existence of field-theoretical
ground states and soliton excitations. Spectral parameters of a
spherical triangle correspond to a statistical mechanics; spectral
parameters of hyperbolic triangles correspond to three different
field theories with massless anisotropic dispersion relations.
\end{abstract}



\maketitle

\section*{Introduction}

The ``circular-'' or ``conic net'' (or discrete orthogonal net)
equations for angular data \cite{BobenkoPinkall,Doliwa-circ,KS98}
take a selected place among all the classical integrable systems
\cite{BoSurBook} on cubic lattice with AKP-type hierarchy.
Algebraically, these equations arise as a Hamiltonian form of
discrete three-wave system \cite{DoliwaSantini,Melbourne}. The
``conic net'' equations are classical $q\to 1$ limit of quantum
``$q$-oscillator'' model \cite{BS05} -- the top of a pyramid of
three-dimensional quantum models -- what guarantees in classics
the existence of Lagrangian density, energy/action and variational
principle \cite{circular}. The existence of quantum counterpart is
an evident advantage of discrete space-time models with respect to
their continuous space-time predecessors
\cite{Zakharov:1973jetp,SolitonBook}.  A straightforward
geometrical condition for conic net equations is the reality of
angular dynamical variables \cite{circular} of a circular net in
Euclidean target space or of an ortho-chronous hyperbolic net in
Minkowski one\footnote{Euclidean sphere of Miquel's theorem
corresponds to a one-sheet hyperboloid in Minkowski metric.}.
However, this discrete differential geometry conditions can be
essentially extended by a ``physical'' condition of reality of
action near equilibrium point.

A general complex solution of any AKP-type system in finite volume
or a solution of Cauchy problem with generic initial data involves
the algebraic geometry \cite{KricheverNovikov}. Such general
solution of discrete ``generalized conic net'' equations in finite
volume is known for a long time \cite{Korepanov:1994lomi1,
Korepanov:1994lomi2, Korepanov:1994lomi3, Korepanov:1994lomi4,
Korepanov:1995, Korepanov:1999tmp}. It involves a flat algebraic
spectral curve $\Gamma_g$ of genus $g\leq(N-1)^2$ for a size
$N^{\times 3}$ cubic lattice (three-periodic boundary conditions),
$\Theta$-functions on Jacobian of $\Gamma_q$, and spectral
parameters -- three meromorphic functions on $\Gamma_g$. A
reduction of $\Gamma_g$ to a sphere gives a $g$-soliton solution
(plane wave solitons) \cite{Q-Toda,Sergeev:PN}. The soliton regime
is the field-theoretical one since solitons are continuous
excitations over a ground state -- zero soliton homogeneous
solution of equations of motion. Spectral parameters in the
soliton regime are a triple of complex numbers, they enter
directly into equations of motion as extra parameters providing
the existence of homogeneous solution. The spectral parameters
break the straightforward discrete-geometric interpretation of
equations of motion.

These two principles --  the reality of energy/action and
existence of homogeneous solution for ``non-zero'' spectral
parameters -- are starting points for classification of physical
field theories and statistical mechanics for ``generalized conic
net'' equations. There are four distinct regimes of spectral
parameters and corresponding regimes of dynamic fields providing
the reality of action. Parameterizations of ground states have a
structure of cosine theorems for spherical or various hyperbolic
triangles. In the case of spherical triangle the ground state is
the absolute minimum of energy functional and thus it corresponds
to statistical mechanics. In three field-theoretical cases of
hyperbolic triangles a solution of equations of motion provides an
extremum of action; a value of whole action on soliton solution
does not depend on amplitudes of solitons.
Expressions for soliton waves involve projective coordinates of
hyperbolic triangles, a general field-theoretical solution of
equations of motion is a set of soliton-anti soliton pairs
analogues to elementary stationary waves. In low energy-momentum
limit the soliton plane waves have a cone-type (anisotropic
massless) dispersion relation.

All the regimes of spectral parameters and dynamical fields have
manifest counterparts in quantum case. Regimes of fields
correspond to classical limits of different representations of
$q$-oscillator algebra; related regimes of spectral parameters
correspond to either real or unitary quantum $R$-matrices.
However, relations between quantum theories and classical theories
are not straightforward.

This paper is organized as follows. In Section 1 we fix notations
for the  ``conic net'' equations generalized by spectral
parameters. Following \cite{circular}, Lagrangian density is
defined in Section 2. Next, in Section 3 we classify ground
states. Soliton solutions of equations of motion and dispersion
relations are defined in Section 4. In Section 5 we describe
quantum counterparts of our four regimes. Finally, in concluding
Section 6 we discuss roughly a place of finite gap solutions.

\section{Generalized conic net equations.}

Let $\nop$ be a node of a big size simple cubic lattice
\begin{equation}\label{lattice}
\nop=(n_1,n_2,n_3)\;,\quad n_i\in\mathbb{Z}\;.
\end{equation}
Let $\e_1,\e_2,\e_3$ be the unit vectors for the lattice,
\begin{equation}
\e_1=(1,0,0)\;,\quad \e_2=(0,1,0)\;,\quad \e_3=(0,0,1)\;,
\end{equation}
so that $\nop=n_1\e_1+n_2\e_2+n_3\e_3$.
With each $(\nop-\e_i,\nop)$ edge of the cubic lattice we
associate a doublet of dynamical variables
\begin{equation}
\alg_{i,\nop}\;=\;(k_{i,\nop}^{},a^{\pm}_{i,\nop})\;,\quad\textrm{such
that }\quad
k_{i,\nop}^2\stackrel{\textrm{def}}{=}1-a^+_{i,\nop}a^-_{i,\nop}\;.
\end{equation}
Last relation here is in fact the ``conic net'' condition. The
local equations of motion relate the neighbors of every node
$\nop$,
\begin{equation}\label{identification}
\alg_i=\alg_{i,\nop}\;\;\textrm{on}\;\;(\nop-\e_i,\nop)\quad
\textrm{and}\quad
\alg_i'=\alg_{i,\nop+\e_i}\;\;\textrm{on}\;\;(\nop,\nop+\e_i)\;,
\end{equation}
as follows:
\begin{equation}\label{themap}
\begin{array}{lll}
\ds (k_2^{} a_1^{\pm})' &=&\ds u_1^{\pm 1} \left(
k_3^{}a_1^{\pm}+u_2^{\mp 1} k_1^{}a_2^{\pm}
a_3^{\mp}\right)\;,\\
\\
\ds (a_2^\pm)' &=&  \ds a_1^\pm a_3^\pm
-u_2^{\mp 1} k_1^{}k_3^{}a_2^{\pm}\;,\\
\\
\ds (k_2^{}a_3^{\pm})' &=& \ds u_3^{\pm 1} \left(
k_1^{}a_3^{\pm}+u_2^{\mp 1} k_3^{}a_1^{\mp}a_2^{\pm}\right)\;,
\end{array}
\end{equation}
and
\begin{equation}\label{kk}
k_1^{}k_2^{}=k_1'k_2'\;,\quad k_2^{}k_3^{}=k_2'k_3'\;.
\end{equation}
$\mathbb{C}$-valued parameters $u_i$ -- exponents of spectral
parameters -- are the same for all nodes $\nop$.

This classical system can be viewed as an extension of discrete
three-wave equations and conic nets since the last ones correspond
to trivial spectral parameters. The circular net in Euclidean
geometry is described by regime $k^2>0$, real $a^{\pm}$ and
$u_1=-u_2=u_3=1$. The Euclidean ``circular net'' point has a
smooth continuous limit -- the classical three-wave resonant
equations \cite{Zakharov:1973jetp}. The hyperbolic net in
Minkowski geometry is described by regime $k^2<0$, real $a^{\pm}$
and $u_1=u_2=u_3=1$. For non-trivial spectral parameters or for
complex $a^{\pm}$ a geometrical interpretation of equations
(\ref{themap}) in unclear.

However, relaxing the geometric condition, equations
(\ref{themap}) define an evolution in $2+1$ dimensional
space-time. Discrete time is $t=n_1+n_2+n_3$, so that
(\ref{themap}) literally gives the map from time $t$ to time
$t+1$. A straightforward way to introduce space-like coordinates
is to take $n_1=x$ and $n_3=y$ so that
\begin{equation}
\nop=\underbrace{n_1(\e_1-\e_2)}_{\ds x\e_x} +
\underbrace{n_3(\e_3-\e_2)}_{\ds y \e_y} +
\underbrace{(n_1+n_2+n_3)\e_2}_{\ds t \e_t}\;.
\end{equation}
The Cauchy problem is well posed for finite size space-like
surface,
\begin{equation}
x,y\in\mathbb{Z}_N\;,\quad N\gg 1\;,\quad t\in\mathbb{Z}_{\geq
0}\;.
\end{equation}
The evolution corresponds to a relativistic field theory since the
locality of evolution map provides the relativistic casuality.
Note, there is no way to introduce a usual local Hamiltonian for
classical discrete time evolution; for instance a principal object
of corresponding quantum theories is a discrete time Heisenberg
evolution operator.
The framework of statistical mechanics implies $3D$ Dirichlet or
$3D$ periodical boundary conditions for system (\ref{themap}),
\begin{equation}
n_i\in\mathbb{Z}_N\;,\quad i=1,2,3\;,\quad N\gg 1\;.
\end{equation}
The principal difference between a statistical mechanics and a
field theory is that for given reality regime and $3D$ periodical
boundary conditions a statistical mechanics has an unique ground
state minimizing an energy while a field theory with a saddle-type
action has a rich structure of stationary modes.

Equations (\ref{themap}) is a canonical transformation preserving
locally the $q$-oscillator symplectic form \cite{BS05},
\begin{equation}
\sum_{i=1}^3 \; \frac{d\, a_i^+\wedge d\, a_i^-}{k_i^2}\;=\;
\sum_{i=1}^3 \; \frac{d\, a_i^{\prime +}\wedge d\, a_i^{\prime
-}}{k_i^{\prime 2}}\;,
\end{equation}
and thus they have the discrete-type Hamiltonian structure
\cite{circular}:
\begin{equation}\label{Hamiltonian}
\log|k_i'|=\frac{1}{2}v_i'\frac{\partial}{\partial
v_i'}\widetilde{{G}}(v;v')\;,\quad
\log|k_i^{}|=-\frac{1}{2}v_i^{}\frac{\partial}{\partial
v_i^{}}\widetilde{{G}}(v;v')\;,
\end{equation}
where
\begin{equation}\label{v}
v_i\;\stackrel{\textrm{def}}{=}\;\frac{a^+_i}{a^-_i}
\end{equation}
is a useful canonical partner to $k_i$, the variables
$v_j^{},v_j',k_j^{},k_j'$ are related by (\ref{themap}), and
$\widetilde{G}(v;v')$ is a generating function of the map
(\ref{themap}).

It is more convenient to treat $k,k'$ related by (\ref{kk}) as
independent variables and define ${G}(k;k')$ by
\begin{equation}\label{differential}
d{G}(k;k')=\frac{1}{2}\sum_{i=1}^3 \left(\log[v'_i]\; d\log k'_i -
\log[v_i^{}]\; d\log k_i^{}\right)\;.
\end{equation}
Here
\begin{equation}
[v]=v\quad \textrm{if $v$ is positive or unitary and}\quad
[v]=-v\quad \textrm{if $v$ is real negative.}
\end{equation}
Negative $v$ corresponds to regime $k^2>1$. Functions ${G}(k;k')$
and $\widetilde{{G}}(v;v')$ are related by the Legendre transform,
\begin{equation}
\widetilde{{G}}(v;v')=\frac{1}{2}\left(\sum_{i=1}^3
\log[v_i']\log|k_i'|-\log[v_i^{}]\log|k_i^{}|\right)\;-\;{G}(k;k')\;.
\end{equation}
Function ${G}(k;k')$ is preferable since $k_i$ are the field
coordinate-type variables with fixed reality regime
$k^2\in\mathbb{R}$.

The sum of local generating functions over all nodes,
\begin{equation}\label{action}
\mathscr{A}\;\sim\; \sum_{\nop\in\mathbb{Z}^3}
\widetilde{{G}}(v_{i,\nop};v_{i,\nop+\e_i})\;,
\end{equation}
gives an action/energy for whole lattice.  Equations of motion are
the extremum conditions for (\ref{action}) \cite{circular}.

Considering the equations of motion as the quasiclassical limit of
quantum models, we expect two distinct regimes: $k^2<0$ for
modular representation of $q$-oscillators and $k^2>0$ for Fock
space representations (regime of unitary $k$ for cyclic
representations we do not consider here). Also, it is evident from
(\ref{differential}) that $\mathscr{A}$ has two regimes of
reality: either the regime of real $v$ when ${G}$ is manifestly
real or the regime of unitary $v$ when $\ii {G}$ is real. Thus we
expect \emph{a priori} the existence of four distinct regimes:
\begin{equation}
\begin{array}{ll}
\ds \textrm{Regime 1:} \quad k_i^2<0\;,\quad |v_i|=1\;,& \ds
\textrm{Regime 2:} \quad k_i^2<0\;,\quad v_i\in\mathbb{R}\;,\\
\ds \textrm{Regime 3:} \quad k_i^2>0\;,\quad |v_i|=1\;,& \ds
\textrm{Regime 4:} \quad k_i^2>0\;,\quad v_i\in\mathbb{R}\;.
\end{array}
\end{equation}

The presence of external parameters $u_i$ in equations of motion
is of great importance for classification scheme because of the
regime $v_i\in\mathbb{R}$ corresponds to $u_i\in\mathbb{R}$ and
the regime $|v_i|=1$ corresponds to $|u_i|=1$.

Note also a signs symmetry of equations (\ref{themap}): a change
of sign of any \emph{two} parameters of $u_i$ followed by a change
of signs of corresponding $a^{\pm}_{i,\nop}$ on e.g. even edges
($n_i$ in $\nop=(n_1,n_2,n_3)$ -- even) does not change the whole
set of equations of motion. This transformation preserves the
variables $v_{i,\nop}$. Contrary to this, a change of a sign of a
single $u_i$ changes a structure of solution of equations of
motion substantially.

\section{Generating function}

The $(v,k)$-form of local equations of motion (\ref{themap}) is
the following \cite{circular}. Let
\begin{equation}\label{tau}
\begin{array}{l}
\ds \tau_0^{}=\frac{v_2'}{v_1'v_3'}\;u_1^2u_3^2\;,\quad
\tau_1^{}=\frac{v_2^{}}{v_1'v_3^{}}\;\frac{u_1^2}{u_2^2}\;,\quad
\tau_2^{}=\frac{v_2'}{v_1^{}v_3^{}}\;,\quad
\tau_3^{}=\frac{v_2^{}}{v_1^{}v_3'}\;\frac{u_3^2}{u_2^2}\;,\\
\\
\ds \tau_0'=\frac{v_2^{}}{v_1^{}v_3^{}}\;\frac{1}{u_2^2}\;,\quad
\tau_1'=\frac{v_2'}{v_1^{}v_3'}\;u_3^2\;,\quad
\tau_2'=\frac{v_2^{}}{v_1'v_3'}\;\frac{u_1^2u_3^2}{u_2^2}\;,\quad
\tau_3'=\frac{v_2'}{v_1'v_3^{}}\;u_1^2\;,
\end{array}
\end{equation}
so that $\tau_0^{}\tau_0'=\tau_i^{}\tau_i'$ and
\begin{equation}\label{tttt}
\tau_i^{}\tau_j^{}=\tau_k'\tau_l'\;,\quad i,j,k,l=\textrm{any
permutation of}\quad (0,1,2,3)\;.
\end{equation}
Equations (\ref{themap}) provide
\begin{equation}
\begin{array}{ll}
\ds
k_1^2=\frac{(1-\tau_2^{})(1-\tau_3^{})}{(1-\tau_0')(1-\tau_1')}\;,&
\ds k_1^{\prime
2}=\frac{(1-\tau_2')(1-\tau_3')}{(1-\tau_0^{})(1-\tau_1^{})}\\
&\\
\ds
k_2^2=\frac{(1-\tau_0')(1-\tau_2')}{(1-\tau_1^{})(1-\tau_3^{})}\;,&
\ds k_2^{\prime 2}=\frac{(1-\tau_0^{})(1-\tau_2^{})}{(1-\tau_1')(1-\tau_3')}\;,\\&\\
\ds
k_3^2=\frac{(1-\tau_1^{})(1-\tau_2^{})}{(1-\tau_0')(1-\tau_3')}\;,&
\ds k_3^{\prime
2}=\frac{(1-\tau_1')(1-\tau_2')}{(1-\tau_0^{})(1-\tau_3^{})}\;.
\end{array}
\end{equation}
Note, expressions for $k_i^2,k_i^{\prime 2}$ are invariant with
respect to inversion $\tau_j^{}\to\tau_j^{-1}$, $\tau_j'\to
\tau_j^{\prime -1}$. Thus, the differential (\ref{differential})
has the form
\begin{equation}\label{differential2}
\begin{array}{l}
\ds d{G}(k;k')=\frac{1}{4}\sum_{j=0}^3 \left(\log[\tau_j^{}]
d\log(1-\tau_j^{})-\log[\tau_j']d\log(1-\tau_j')\right)\\\\
\ds  + \frac{1}{2}\left( \log u_1^2 d\log k_1' - \log u_2^2 d\log
k_2^{} + \log u_3^2 d\log k_3'\right)\;.
\end{array}
\end{equation}
Integrating this, one gets
\begin{equation}\label{g-answer}
\ds {G}(k;k')={G}_0(k;k')  + \frac{1}{2}\left( \log u_1^2 \log
|k_1'| - \log u_2^2 \log |k_2^{}| + \log u_3^2 \log
|k_3'|\right)\;,
\end{equation}
where
\begin{equation}
{G}_0(k;k')=\frac{1}{4}\sum_{j=0}^3 \left( J(\tau_j^{})-
J(\tau_j')\right)
\end{equation}
and
\begin{equation}
J(\tau)= \left\{\begin{array}{l}\ds  \int_{z_0}^{\tau} \log z \;
d\log(1-z)\quad \textrm{in Regimes 1,2}\\ \\
\ds \int_{z_0}^{-\tau} \log z \; d\log(1+z)\quad \textrm{in
Regimes 3,4.}\end{array}\right.
\end{equation}
Choice of $z_0$ common for all integrals is irrelevant. We choose
zero value for a constant of integration in (\ref{g-answer}). The
$G_0(k;k')$-term in (\ref{g-answer}) corresponds to Lagrangian
density without fields \cite{circular} but it must be treated
carefully due to inversion symmetry of $\tau_j$.

Generation function (\ref{g-answer}) becomes symmetric after an
elementary gauge transformation,
\begin{equation}\label{g-symmetric}
\ds {G}_{sym}(k;k')={G}_0(k;k')   + \frac{1}{4}\left( \log u_1^2
\log |k_1^{}k_1'| - \log u_2^2 \log |k_2^{}k_2'| + \log u_3^2 \log
|k_3^{}k_3'|\right).
\end{equation}

\section{Ground states}

By ``ground state'' we understand the \emph{homogeneous} solutions
of equations (\ref{themap}):
\begin{equation}
v_j^{}=v_j'\;,\quad k_j^{}=k_j'\;,
\end{equation}
what corresponds to
\begin{equation}
\frac{v_2}{v_1v_3}\;=\;\frac{u_2}{u_1u_3}\;\;\Leftrightarrow\;\;\tau_j'=\tau_j^{-1}\;.
\end{equation}
In the vicinity of this point the fields are parameterized by
\begin{equation}\label{kkk}\begin{array}{ll}
\ds
k_1^2=\frac{(u_1u_3-u_2)(u_1u_2-u_3)}{(1-u_1u_2u_3)(u_2u_3-u_1)}\EXP^{2\rho_1^{}}\;,&
\ds k_1^{\prime
2}=\frac{(u_1u_3-u_2)(u_1u_2-u_3)}{(1-u_1u_2u_3)(u_2u_3-u_1)}\EXP^{2\rho_1^{\prime}}\;,
\\\\
\ds
k_2^2=\frac{(1-u_1u_2u_3)(u_1u_3-u_2)}{(u_2u_3-u_1)(u_1u_2-u_3)}\EXP^{2\rho_2^{}}\;,&
\ds k_2^{\prime
2}=\frac{(1-u_1u_2u_3)(u_1u_3-u_2)}{(u_2u_3-u_1)(u_1u_2-u_3)}\EXP^{2\rho_2^{\prime}}\;,
\\\\
\ds
k_3^2=\frac{(u_2u_3-u_1)(u_1u_3-u_2)}{(1-u_1u_2u_3)(u_1u_2-u_3)}\EXP^{2\rho_3^{}}\;,
& \ds k_3^{\prime
2}=\frac{(u_2u_3-u_1)(u_1u_3-u_2)}{(1-u_1u_2u_3)(u_1u_2-u_3)}\EXP^{2\rho_3^{\prime}}\;,
\end{array}
\end{equation}
where
\begin{equation}
\rho_1^{}-\rho_1'=\rho_2'-\rho_2^{}=\rho_3^{}-\rho_3'\;.
\end{equation}
The generating function (\ref{g-symmetric}) near the ground state
$\rho=0$ is
\begin{equation}\label{free-field}
\begin{array}{l}
\ds G_{sym}\;=\;G_{sym}|_{\rho=\rho'=0}+ \sum_{i}
\frac{(u_i^{}-u_i^{-1})}{(u_j^{}-u_j^{-1})(u_k^{}-u_k^{-1})}\; x_i^2\\
\\
\ds -\frac{u_1^{}+u_1^{-1}}{u_1^{}-u_1^{-1}}\; x_2^{}x_3^{}
+\frac{u_2^{}+u_2^{-1}}{u_2^{}-u_2^{-1}}\; x_1^{}x_3^{}
-\frac{u_3^{}+u_3^{-1}}{u_3^{}-u_3^{-1}}\; x_1^{}x_2^{}\\
\\
\ds
-\frac{(u_1u_2u_3-1)(u_1u_2-u_3)(u_1u_3-u_2)(u_2u_3-u_1)}{u_1^2u_2^2u_3^2(u_1^{}-u_1^{-1})
(u_2^{}-u_2^{-1})(u_3^{}-u_3^{-1})}\;\delta^2\;+\;\mathcal{O}(\rho^3)\;,
\end{array}
\end{equation}
where
\begin{equation}
x_i=\frac{\rho_i^{}+\rho_i'}{2}\;,\quad
\delta^2=\left(\frac{\rho_i^{}-\rho_i'}{2}\right)^2\;,
\end{equation}
and $G_{sym}|_{\rho=\rho'=0}=G(k;k)$ is calculated at the point
$\rho_i^{}=\rho_i'=0$ ($k_i^{}=k_i'$).

Now we are ready to classify the generating functions for all four
regimes. Depending on regime, expressions (\ref{kkk}) are
equivalent to cosine theorems for spherical or hyperbolic
triangles; thus the classification scheme is based on the
spherical and hyperbolic geometry.

\subsection{Regime 1} Unitary spectral parameters are given by
\begin{equation}\label{u-reg1}
u_1=\EXP^{\ii\epsilon_1\phi_1}\;,\quad
u_2=\EXP^{\ii\epsilon_2\phi_2}\;,\quad
u_3=\EXP^{\ii\epsilon_3\phi_3}\;,
\end{equation}
where $\phi_i>0$ and $\epsilon_i$ are signs. In what follows we
use short notations
\begin{equation}\label{excess}
\beta_4=\frac{\phi_1+\phi_2+\phi_3}{2}\;,\quad
\beta_i=\beta_4-\phi_i\;.
\end{equation}
In spherical geometry the angles $\phi_i$ are sides of a spherical
triangle. It is convenient to define excess $\beta_0$ instead of
half-perimeter $\beta_4$ by
\begin{equation}\label{beta0}
\beta_0=\pi-\beta_4\;.
\end{equation}
Then for $\epsilon_1=\epsilon_2=\epsilon_3=1$ one has
\begin{equation}\label{k-tan}
k_1^2=-\tan^2\frac{\theta_1}{2}\;,\quad
k_2^2=-\cot^2\frac{\theta_2}{2}\;,\quad
k_3^2=-\tan^2\frac{\theta_3}{2}\;,
\end{equation}
where $\theta_i$ are dihedral angles of the spherical triangle
with sides $\phi_i$:
\begin{equation}
\begin{array}{l}
\ds
\cos\theta_i=\frac{\cos\phi_i-\cos\phi_j\cos\phi_k}{\sin\phi_j\sin\phi_k}\;,\quad
\cos\phi_i=\frac{\cos\theta_i+\cos\theta_j\cos\theta_k}{\sin\theta_j\sin\theta_k}\;,\\
\\
\ds \tan^2\frac{\theta_i}{2}=\frac{\sin\beta_j\sin\beta_k}{\sin
\beta_0 \sin\beta_i}\;.
\end{array}
\end{equation}
Spherical geometry implies positive $\theta$-excess
\begin{equation}
\theta_1+\theta_2+\theta_3>\pi\;\;\Rightarrow\;\;
0<\beta_j<\pi\;,\;\;j=0,1,2,3\;.
\end{equation}
Other choices of signs $\epsilon_i$ such that
$\epsilon_1\epsilon_2\epsilon_3=1$ are equivalent to crossing
transformations of spherical triangle (in fact, crossing
transformations involve $\phi\to\pi-\phi$, the sign symmetry of
$u_i$ we discussed above).

For arbitrary signs $\epsilon_i$ let
\begin{equation}\label{Energy}
F(k;k')\;=\;\ii\epsilon_1\epsilon_2\epsilon_3{G}(k;k')\;,\quad
\mathscr{H}\;=\;\sum_{\nop\in\mathbb{Z}^3}
F(k_{i,\nop};k_{i,\nop+\e_i})\;.
\end{equation}
Homogeneous solution provides the absolute minimum of functional
$\mathscr{H}$; on this ground state the free energy density
$F(k;k)=F_0$ is given by
\begin{equation}\label{free-energy}
F_0=\sum_{j=0}^3 \Lb(\beta_j)\;=\;\sum_{i=1}^3
\Lb(\beta_i)-\Lb(\beta_4)\;>\;0\;,
\end{equation}
where the Milnor's Lobachevski function is
\begin{equation}
\Lb(\beta)=-\int_0^\beta \log(2\sin x) \, dx\;.
\end{equation}
The statement about absolute minimum can be verified instantly in
the free field approximation (\ref{free-field}) where
\begin{equation}
\mathscr{H}\;=\;N^3 F_0+\textrm{positively defined quadratic form
of $\rho_{i,\nop}$}\;.
\end{equation}
Here $N^3$ is a volume of the lattice. We will discuss this
statement beyond the free field approximation in the next section.

\subsection{Regime 2} This is the case of spectral parameters
\begin{equation}\label{u-reg2}
u_1=\EXP^{\epsilon_1\phi_1}\;,\quad
u_2=\EXP^{\epsilon_2\phi_2}\;,\quad u_3=\EXP^{\epsilon_3\phi_3}\;,
\end{equation}
where $\phi_i>0$ and $\epsilon_i$ are signs. Values of $k_i$ of
homogeneous solution for $\epsilon_1=\epsilon_2=\epsilon_3=1$ are
given by (\ref{k-tan}) where $\theta_i$ are dihedral angles of a
triangle on upper sheet of two-sheets hyperboloid. Parameters
$\phi_i$ are hyperbolic sides of this triangle. Cosine theorems
read
\begin{equation}
\begin{array}{l}
\ds
\cos\theta_i=\frac{\cosh\phi_j\cosh\phi_k-\cosh\phi_i}{\sinh\phi_j\sinh\phi_k}\;,\quad
\cosh\phi_i=\frac{\cos\theta_i+\cos\theta_j\cos\theta_k}{\sin\theta_j\sin\theta_k}\;,\\
\\
\ds
\tan^2\frac{\theta_i}{2}=\frac{\sinh\beta_j\sinh\beta_k}{\sinh\beta_4
\sinh\beta_i}\;,
\end{array}
\end{equation}
where excesses are given by (\ref{excess}). Hyperbolic geometry
implies negative $\theta$-excess
\begin{equation}
\theta_1+\theta_2+\theta_3<\pi\quad\Rightarrow\quad
0<\beta_i\;,\;\;i=1,2,3\;.
\end{equation}
Other choices of signs $\epsilon_i$ are analogues of crossing
transformation of the hyperbolic triangle.

For arbitrary signs $\epsilon_i$ define the Lagrangian density and
the action by
\begin{equation}
L(k;k')\;=\;-\epsilon_1\epsilon_2\epsilon_3 {G}(k;k')\;,\quad
\mathscr{A}\;=\;\sum_{\nop\in\mathbb{Z}^3}
L(k_{i,\nop};k_{i,\nop+\e_i})\;.
\end{equation}
The criterion for a correct sign of Lagrangian density is the
positive sign near $\delta^2$ in the free-field approximation
(\ref{free-field}),
\begin{equation}\label{canonical-Lagrangian}
L(k;k')= (\textrm{positive coeff.}) \times \delta^2 - V(x) -
V_0\;,
\end{equation}
so that $\delta^2$ stands for a square of velocity and $V(x)+V_0$
stands for a potential. In this regime the quadratic form $V(x)$
is positively defined. On homogeneous solution (ground state)
\begin{equation}
L(k;k)=-V_0=\int_0^{\beta_4} \log(2\sinh x)\,dx - \sum_{i=1}^3
\int_0^{\beta_i} \log(2\sinh x)\,dx\;>\;0\;.
\end{equation}
The global quadratic form has saddle structure, general solution
of linearized equations of motion are plane waves with a certain
dispersion relation. As it is clear for free theory, the value of
the whole action on any plane wave solution of equations of motion
in finite volume coincides with its value on vacuum solution,
\begin{equation}\label{whole-action}
\mathcal{A}\;=\;-N^3\;V_0\;.
\end{equation}

In the next sections we give a general solitonic solution of
field-theoretical equations of motion which can be regarded as
excitations over the ground state. Dispersion relation for
solitons is the same as the dispersion relation for linearized
theory. Since for $3D$ periodical boundary conditions the value of
whole action at equilibrium point is an universal invariant, it
does not depend on amplitudes of solitons and therefore it equals
to value of the whole action for ground state.

\subsection{Regime 3} Unitary parameters $u_i$ are given by
\begin{equation}\label{u-reg3}
u_1=\EXP^{\ii\epsilon_1\phi_1}\;,\quad
u_2=-\EXP^{\ii\epsilon_2\phi_2}\;,\quad
u_3=\EXP^{\ii\epsilon_3\phi_3}
\end{equation}
where as usual $\phi_i>0$, $\epsilon_i$ are the signs and the
excesses are defined by (\ref{excess}). This gives for
$\epsilon_1=\epsilon_2=\epsilon_3=1$
\begin{equation}\label{k-reg3}
k_1^2=\coth^2\frac{\theta_1}{2}\;,\quad
k_2^2=\tanh^2\frac{\theta_2}{2}\;,\quad
k_3^2=\coth^2\frac{\theta_3}{2}\;,
\end{equation}
with cosine theorems
\begin{equation}
\begin{array}{l}
\ds
\cosh\theta_i=\frac{\cos\phi_i+\cos\phi_j\cos\phi_k}{\sin\phi_j\sin\phi_k}\;,\quad
\cos\phi_i=\frac{\cosh\theta_j\cosh\theta_k-\cosh\theta_i}{\sinh\theta_j\sinh\theta_k}\;,\\
\\
\ds
\coth^2\frac{\theta_i}{2}=\frac{\cos\beta_j\cos\beta_k}{\cos\beta_4
\cos\beta_i}\;.
\end{array}
\end{equation}
This is a hyperbolic triangle formed by an intersection of three
planes with time-like normals (and hyperbolic angles $\theta_i$
between them) and one-sheet hyperboloid. Trigonometric sides
$\phi_i$ are defined in motionless frame of reference for each
plane. The time-like normals form a dual triangle on two-sheets
hyperboloid of Regime 2. The geometry provides the constraint for
$\beta_i$:
\begin{equation}
0<\beta_i<\pi/2\;,\quad i=1,2,3,4\;,
\end{equation}
otherwise it would be Regime 1. Other choices of signs
$\epsilon_i$ are analogues of crossing transformation of the
hyperbolic triangle.

We define the Lagrangian density for arbitrary signs $\epsilon_i$
by
\begin{equation}
L(k;k')=\ii\epsilon_1\epsilon_2\epsilon_3{G}(k;k')\;,
\end{equation}
where the sign criterion is the same as for Regime 2. However, the
quadratic potential here is not sign defined. On homogeneous
solution (ground state)
\begin{equation}
L(k;k)=-V_0=\int_0^{\beta_4} \log(2\cos x)\,dx-\sum_{i=1}^3
\int_0^{\beta_i} \log(2\cos x)\,dx\;<\;0\;.
\end{equation}

\subsection{Regime 4} Real spectral parameters are defined by
\begin{equation}\label{u-reg4}
u_1=\EXP^{-\epsilon_1\phi_1}\;,\quad
u_2=-\EXP^{-\epsilon_2\phi_2}\;,\quad
u_3=\EXP^{-\epsilon_3\phi_3}\;,
\end{equation}
where $\phi_i$ are positive, $\epsilon_i$ are again signs.
Negative sign of one of $u_i$ makes the difference with Regime 2.
This parameterization gives for
$\epsilon_1=\epsilon_2=\epsilon_3=1$
\begin{equation}\label{k-reg4}
k_1^2=\tanh^2\frac{\theta_1}{2}\;,\quad
k_2^2=\coth^2\frac{\theta_2}{2}\;,\quad
k_3^2=\tanh^2\frac{\theta_3}{2}
\end{equation}
where cosine theorems are
\begin{equation}\label{cos-reg4}
\begin{array}{l}
\ds
\cosh\theta_i=\frac{\cosh\phi_i+\cosh\phi_j\cosh\phi_k}{\sinh\phi_j\sinh\phi_k}\;,\quad
\cosh\phi_i=\frac{\cosh\theta_i+\cosh\theta_j\cosh\theta_k}{\sinh\theta_j\sinh\theta_k}\;,\\
\\
\ds
\tanh^2\frac{\theta_i}{2}=\frac{\cosh\beta_j\cosh\beta_k}{\cosh\beta_4
\cosh\beta_i}\;.
\end{array}
\end{equation}
The hyperbolic excesses are defined by (\ref{excess}). This
corresponds to a triangle on one-sheet hyperboloid with hyperbolic
sides $\phi_i$ and hyperbolic dihedral angles $\theta_i$. Such
triangle is the section of one-sheet hyperboloid by planes with
space-like normals. Note, two planes with space-like normals and
hyperbolic angles between them do not intersect on  two-sheet
hyperboloid. Contrary, two planes with space-like normals and
trigonometric angle between them intersect on two-sheets
hyperboloid, this corresponds to Regime 2.

Other choices of signs $\epsilon_i$ are analogues of crossing
transformation of the hyperbolic triangle.

Lagrangian density and the action are then
\begin{equation}
L(k;k')=\epsilon_1\epsilon_2\epsilon_3{G}(k;k')\;,\quad
\mathscr{A}\;=\;\sum_{\nop\in\mathbb{Z}^3}
L(k_{i,\nop};k_{i,\nop+\e_i})\;.
\end{equation}
In this regime the quadratic potential is not sign defined either.
Ground state gives
\begin{equation}\label{V-reg4}
L(k;k)=-V_0\;=\;\int_0^{\beta_4} \log(2\cosh x)\, dx-\sum_{j=1}^3
\int_0^{\beta_j} \log(2\cosh x)\, dx\;>0\;.
\end{equation}

\section{Solitons}

\subsection{General soliton solution of (\ref{themap})}

A general (complex) soliton solution of equations (\ref{themap})
is given by the reduction of general algebraic geometry solution
corresponding to the reduction of genus $g$ curve to a sphere with
punches. Resulting expressions are the following \cite{Q-Toda}.
For the number of solitons $g\geq 0$ let
\begin{equation}
\{X_j,Y_j\}_{j=1,...,g}\;,\quad \boldsymbol{f}=\{f_j\}_{j=1..g}
\end{equation}
be a set of $3g$ complex values. For given $\boldsymbol{f}$ and
$\{X_j,Y_j\}$ let
\begin{equation}
F_j\;=\;f_j\prod_{k\neq j} \frac{X_j-X_k}{Y_j-X_k}\;.
\end{equation}
Define next
\begin{equation}\label{Theta}
\Theta(\boldsymbol{f})=\frac{\ds
\det|X_j^{k-1}+F_j^{}Y_j^{k-1}|_{j,k=1..g}}{\prod_{i>j}(X_i-X_j)}\;,
\end{equation}
where in the numerator there is the determinant of $g\times g$
matrix with matrix indices $j,k$. In this expression $f_j$ is the
amplitude of $j$th soliton, if one of $f_j=0$ then (\ref{Theta})
simply gives $g-1$ soliton expression. For instance,
\begin{equation}
\begin{array}{l}
\ds g=0\quad\Rightarrow\quad \Theta=1\;,\\
\\
\ds g=1\quad\Rightarrow\quad \Theta(f_1)=1+f_1\;,\\
\\
\ds g=2\quad\Rightarrow\quad
\Theta(f_1,f_2)=1+f_1+f_2+f_1f_2\frac{(X_1-X_2)(Y_1-Y_2)}{(X_1-Y_2)(Y_1-X_2)}\;,
\end{array}
\end{equation}
etc. In general, at the first order of $f$
\begin{equation}
\Theta(\boldsymbol{f})\;=\;1+\sum_{j=1}^g f_j + \textrm{higher
terms,}
\end{equation}
what corresponds to the free field (linear) approximation. Let
further
\begin{equation}\label{frequency}
\omega_{k,j}=\omega_k(X_j,Y_j)\;=\;\frac{(Y_j-P_k)(X_j-Q_k)}{(X_j-P_k)(Y_j-Q_k)}\;,\quad
j=1,...,g,\quad k=1,2,3
\end{equation}
and
\begin{equation}\label{plane-wave}
\boldsymbol{f}(\nop)\;=\; \{f_j(\nop)\}_{j=1,...,g}\;,\quad
f_j^{}(\nop)\;=\;f_j^{}\,\omega_{1,j}^{n_1}\omega_{2,j}^{-n_2}\omega_{3,j}^{n_3}\;.
\end{equation}
Let also for brevity
\begin{equation}
\Theta_{\nop}=\Theta(\boldsymbol{f}(\nop))\;.
\end{equation}
General soliton solution of (\ref{themap}) is then given by
\cite{Korepanov:1995}
\begin{equation}\label{k-alg}
\begin{array}{l}
\ds
k_{1,\nop}^2\;=\;\frac{E(Q_2,Q_3)E(P_2,P_3)}{E(Q_2,P_3)E(P_2,Q_3)}
\;\frac{\Theta_{\nop}\Theta_{\nop-\e_2+\e_3}}{\Theta_{\nop-\e_2}\Theta_{\nop+\e_3}}\;,\\
\\
\ds
k_{2,\nop}^2\;=\;\frac{E(Q_1,Q_3)E(P_1,P_3)}{E(Q_1,P_3)E(P_1,Q_3)}
\;\frac{\Theta_{\nop-\e_2}\Theta_{\nop+\e_1-\e_2+\e_3}}{\Theta_{\nop+\e_1-\e_2}\Theta_{\nop-\e_2+\e_3}}\;,\\
\\
\ds
k_{3,\nop}^2\;=\;\frac{E(Q_1,Q_2)E(P_1,P_2)}{E(Q_1,P_2)E(P_1,Q_2)}
\;\frac{\Theta_{\nop}\Theta_{\nop+\e_1-\e_2}}{\Theta_{\nop+\e_1}\Theta_{\nop-\e_2}}\;,
\end{array}
\end{equation}
where
\begin{equation}
E(Q,P)=\frac{Q-P}{\sqrt{dQdP}}
\end{equation}
is the prime form on a compact complex plane. Spectral parameters
in this parameterization are given by
\begin{equation}\label{u-via-E}
\begin{array}{ll}
u_1u_2u_3=-\frac{E(Q_1,Q_2)E(P_1,Q_3)E(P_2,P_3)}{E(P_1,P_2)E(Q_1,P_3)E(Q_2,Q_3)}\;,&
\frac{u_1}{u_2u_3}=-\frac{E(Q_1,P_2)E(P_1,P_3)E(Q_2,Q_3)}{E(P_1,Q_2)E(Q_1,Q_3)E(P_2,P_3)}\;,\\&\\
\frac{u_2}{u_1u_2}=-\frac{E(P_1,Q_2)E(Q_1,P_3)E(P_2,Q_3)}{E(Q_1,P_2)E(P_1,Q_3)E(Q_2,P_3)}\;,&
\frac{u_3}{u_1u_2}=-\frac{E(P_1,P_2)E(Q_1,Q_3)E(Q_2,P_3)}{E(Q_1,Q_2)E(P_1,P_3)E(P_2,Q_3)}\;.
\end{array}
\end{equation}

\subsection{Identification of parameterizations}

Homogeneous solution of (\ref{themap}) corresponds to $g=0$ when
all $\Theta_{\nop}=1$. Expressions (\ref{k-alg}) are equivalent to
parameterizations of homogeneous $k_j^2$ in terms of spherical and
hyperbolic triangles. Let us demonstrate this statement in more
details.

In Regime 1 of Euclidean spherical trigonometry, consider thee
planes defined by their unit normal vectors $\vec{n}_i$ in some
auxiliary frame of reference,
\begin{equation}
\vec{n}_i=(\sin\vartheta_i\cos\varphi_i,\sin\vartheta_i\sin\varphi_i,\cos\vartheta_i)
\end{equation}
The dihedral angle between two planes equals to the angle between
normals,
\begin{equation}
\theta_3=\widehat{\vec{n}_1\vec{n_2}}\;,\quad
\theta_2=\pi-\widehat{\vec{n}_1\vec{n}_3}\;,\quad \theta_1=
\widehat{\vec{n}_2\vec{n}_3}\;,
\end{equation}
where $\theta_i$ are \emph{inner} dihedral angles of the spherical
triangle. Cosine theorems give
\begin{equation}
\cos\theta_1=(\vec{n}_2,\vec{n}_3)=\cos\vartheta_2\cos\vartheta_3+\sin\vartheta_2\sin\vartheta_3\cos(\varphi_2-\varphi_3)
\quad \textrm{etc.}
\end{equation}
For all points $\vec{n}_i$ on the sphere define their
stereographic projections to a complex plane:
\begin{equation}
Q_i\;=\;\tan\frac{\vartheta_i}{2}\EXP^{\ii\varphi_i}\;,\quad
P_i=-\cot\frac{\vartheta_i}{2}\EXP^{\ii\varphi_i}\;.
\end{equation}
The cosine theorem can be then rewritten as
\begin{equation}
-\tan^2\frac{\theta_1}{2}\;=\;\frac{\cos\theta_1-1}{\cos\theta_1+1}\;=\;\frac{(Q_2-Q_3)(P_2-P_3)}{(Q_2-P_3)(P_2-Q_3)}\;,
\end{equation}
what makes exact correspondence between (\ref{k-tan}) and
(\ref{k-alg}) for $\Theta_{\nop}=1$.

This can be done similarly for all other regimes. In Minkowski
metric $\boldsymbol{g}=\textrm{diag}(1,-1,-1)$, a time-like unit
vector is parameterized by
\begin{equation}
\begin{array}{l}
\ds
\vec{n}_i=(\cosh\vartheta_i,\sinh\vartheta_i\cos\varphi_i,\sinh\vartheta_i\sin\varphi_i)\;,
\quad \textrm{so that}\\
\\
\ds  \cosh\theta_1=(\vec{n}_2,\vec{n_3})=
\cosh\vartheta_1\cosh\vartheta_2-\sinh\vartheta_1\sinh\vartheta_2\cos(\varphi_1-\varphi_2)\;,
\end{array}
\end{equation}
what gives for Regime 3
\begin{equation}
Q_i=\tanh\frac{\vartheta_i}{2}\EXP^{\ii\varphi_i}\;,\quad
P_i=\coth\frac{\vartheta_i}{2}\EXP^{\ii\varphi_i}\;\;.
\end{equation}
The proper parameterization of space-like unit vectors is the
following:
\begin{equation}
\begin{array}{l}
\ds
\vec{n}_i=(\tan\vartheta_i,\sec\vartheta_i\cos\varphi_i,\sec\vartheta_i\sin\varphi_i)\;,
\quad \textrm{so that}\\
\\
\ds
\cos\theta_1\;\;\textrm{or}\;\;\cosh\theta_1=-(\vec{n}_2,\vec{n_3})=
\frac{\cos(\varphi_2-\varphi_3)-\sin\vartheta_1\sin\vartheta_2}{\cos\vartheta_2\cos\vartheta_3}\;,
\end{array}
\end{equation} what gives for Regimes 2,4
\begin{equation}
Q_i=\EXP^{\ii(\varphi_i+\vartheta_i)}\;,\quad
P_i=-\EXP^{\ii(\varphi_i-\vartheta_i)}\;\;.
\end{equation}

Thus, in terms of complex parameters $P_i,Q_i$, Regimes are
classified as follows:
\begin{equation}\label{regimes-PQ}
\begin{array}{ll}
\ds\textrm{Regime 1:}& \ds P_i^{}Q_i^*=-1\;,\\
\ds\textrm{Regime 3:} & \ds P_i^{}Q_i^*=1\;,\\
\ds\textrm{Regimes 2,4:} & \ds |P_i|=|Q_i|=1\;,\\
\end{array}
\end{equation}
where $\phantom{|}^*$ stands for complex conjugation.

\subsection{Plane waves and dispersion relation}

Relation (\ref{plane-wave}) stands for a plane wave with
exponential frequencies $\omega_k(X,Y)$, $k=1,2,3$.
Parameterization (\ref{frequency}) can be viewed as a general
solution of an algebraic equation relating three $\omega_k$,
$k=1,2,3$. This dispersion relation can be obtained by elimination
of $X,Y$ from (\ref{frequency}), it has the form
\begin{equation}
\sum_{i,j,k=0}^2
c_{ijk}(u_1,u_2,u_3)\omega_1^i\omega_2^j\omega_3^k=0\;,
\end{equation}
where $c_{i,j,k}(u_1,u_2,u_3)$ are simple but lengthy rational
coefficients. Note that the free field approximation provides the
same dispersion relation.

Three-periodical boundary conditions in a rather big volume
require unitary $\omega_k$. Parameterization (\ref{frequency}) and
definition of regimes (\ref{regimes-PQ}) provide immediately the
following:
\begin{itemize}
\item In Regime 1 it is impossible\footnote{Unitarity condition in Regime 1 demands $X^*X=Y^*Y=-1$.}
to make all three $\omega_{k}(X,Y)$
unitary. Thus, the homogeneous solution is indeed the absolute
minimum of the energy functional (\ref{Energy}). For the open
boundary conditions, the solitons of Regime 1 break the signature
condition $k_{i,\nop}^2<0$.
\item In Regimes 2,4, when $P_k,Q_k$ are unitary, all  $\omega_{k}(X,Y)$  are
unitary if $X^*Y^{}=1$.
\item
In Regime 3, when $P_k^{}Q_k^*=1$, all $\omega_{k}(X,Y)$  are
unitary if $|X^{}|=|Y^{}|=1$.
\end{itemize}
In all field-theoretical regimes the reality condition
$\Theta_{\nop}^*=\Theta_{\nop}^{}$ is satisfied for
soliton-antisoliton pairs with conjugated amplitudes.

The dispersion relation for $\omega_k$ near unity,
\begin{equation}
\omega_1=\EXP^{\ii p_1}\;,\quad \omega_2=\EXP^{-\ii p_2}\;,\quad
\omega_3=\EXP^{\ii p_3}\;,
\end{equation}
where momenta $p_i$ are small, reads
\begin{equation}\label{ds-4}
p_1^2p_2^2(u_3^{}-u_3^{-1})^2+p_1^{}p_2^{}p_3^2(u_1^{}-u_1^{-1})(u_2^{}-u_2^{-1})(u_3^{}+u_3^{-1})
+\textrm{cyclic permutations}=0.
\end{equation}
Due to the homogeneouty, this relation describes a cone-type
surface in momentum space.
In the symmetric cases $u_1=u_2=u_3$ or $u_1=-u_2=u_3$, what
corresponds to $\phi_1=\phi_2=\phi_3=\phi>0$, relation
(\ref{ds-4}) becomes
\begin{equation}\label{kone}
p_1^2p_2^2+p_1^2p_3^2+p_2^2p_3^2\;+\;2\, C\,
p_1^{}p_2^{}p_3^{}(p_1^{}+p_2^{}+p_3^{})=0\;,
\end{equation}
where
\begin{equation}\label{regimes-c}
\begin{array}{l}
\ds \textrm{Regime 1:}\quad C=\cos\phi,\;,\quad -1/2<C<1\;;\\
\ds \textrm{Regime 2:}\quad C=\cosh\phi\;,\quad 1<C\;;\\
\ds \textrm{Regime 3:}\quad C=-\cos\phi\;,\quad -1<C<-1/2\;; \\
\ds \textrm{Regime 4:}\quad C=-\cosh\phi\;,\quad
C<-1\;.\end{array}
\end{equation}
Our four regimes cover the real axis,
$C\in\mathbb{R}\backslash\{1,-1,-1/2\}$. Define then an energy $E$
and space-like momenta $\pi_i$ (we do not care about scales of
energy and momenta) by
\begin{equation}
p_i=E+\pi_i\;,\quad \pi_1+\pi_2+\pi_3=0\;.
\end{equation}
Let
\begin{equation}
\pi^2=\frac{1}{2}(\pi_1^2+\pi_2^2+\pi_3^2)\;,\quad
\gamma=\frac{\pi_1\pi_2\pi_3}{\pi^3}\;.
\end{equation}
If $C<-1/2$ or $1<C$, equation (\ref{kone}) defines an anisotropic
cone-type surface
\begin{equation}
E=\alpha(\gamma)\pi
\end{equation}
where $\alpha(\gamma)$ is a real solution of
\begin{equation}\label{quadric}
(C+\frac{1}{2})\alpha^4-C\alpha^2+(C-1)\gamma\alpha+\frac{1}{6}=0\;.
\end{equation}
Anisotropy parameter $\gamma$ is bounded,
\begin{equation}
-\gamma_0\leq\gamma\leq\gamma_0\;,\quad
\gamma_0=\sqrt{\frac{4}{27}}\;.
\end{equation}
Critical values $\gamma=\pm\gamma_0$ correspond to three selected
directions in the momentum space when
\begin{equation}
p_1=p_2=0\quad \textrm{or}\quad p_1=p_3=0\quad \textrm{or}\quad
p_2=p_3=0\;.
\end{equation}
When $C<-1/2$ (Regimes 3,4), equation (\ref{quadric}) has one
positive and one negative solution what gives a rather anisotropic
``cone'' with
\begin{equation}
\frac{1}{\sqrt{3}}\left(\sqrt{\frac{2C-2}{2C+1}}-1\right)\leq
\alpha_+(\gamma)\leq
\frac{1}{\sqrt{3}}\left(\sqrt{\frac{2C-2}{2C+1}}+1\right)\;,
\end{equation}
where $\alpha_+(\gamma)$ are taken positive (negative solutions
for given $\gamma$ are $\alpha_-(\gamma)=-\alpha_+(-\gamma)$).
When $\gamma=\pm\gamma_0$, equation (\ref{quadric}) has extra
solutions $\alpha=\pm \frac{1}{\sqrt{3}}$, these solutions are
isolated and therefore do not belong to one-parameter family, they
have no relation to $\omega_i$ and should be ignored.

When $1<C$ (Regime 2), equation (\ref{quadric}) has two positive
and two negative solutions what gives two imbedded tangent
anisotropic ``cones''. The ``cones'' are tangent along
$\gamma=\pm\gamma_0$, $\alpha(\pm\gamma_0)=\pm
\frac{1}{\sqrt{3}}$, these points are not isolated. Existence of
two ``speeds of lights'' is a surprise.

Regime 2 involves the Lorentz group limit. If all $\phi_i$ in this
Regime are small, $\cos\phi_1\simeq 1$, then the dispersion
relation (\ref{ds-4}) becomes
\begin{equation}
\left( \frac{p_1}{\phi_1}\,\frac{p_2}{\phi_2} \, + \,
\frac{p_1}{\phi_1}\,\frac{p_3}{\phi_3} \, + \,
\frac{p_2}{\phi_2}\,\frac{p_3}{\phi_3} \right)^2\;=\;0\;,
\end{equation}
what is equivalent to isotropic light cone and gives the pure
Minkowski metric in the momentum space.

When $-1/2<C<1$ (Regime 1), equation (\ref{quadric}) has no real
solutions as expected.

\section{Quantum theories}

In this section we discuss the relation of classical regimes to
quantum models.

We commence with a short remainder of quantum $R$-matrices. Let
$\alg$ be the enveloping of the $q$-oscillator algebra
\begin{equation}\label{qosc}
\kop \bos^{\pm} = q^{\pm 1} \bos^{\pm} \kop\;,\quad
\bos^{+}\bos^{-}=1-q^{-1} \kop^2\;,\quad \bos^-\bos^+=1-q\kop^2\;.
\end{equation}
equipped by a pair of $\mathbb{C}$-valued parameters
$\lambda,\mu$,
\begin{equation}
\alg=(1,\kop,\bos^{\pm},\lambda,\mu)\;.
\end{equation}
The map $\Rop_{123}$ \cite{BS05,circular} of tensor cube
$\alg_1\otimes \alg_2\otimes \alg_3 $ defined by (confer with
(\ref{themap}))
\begin{equation}\label{Themap}
\begin{array}{lll}
\ds \Rop_{123}^{} \;\kop_2^{}\bos_1^{\pm}\; \Rop_{123}^{-1} &=&
\ds u_1^{\pm 1} \left( \kop_3^{}\bos_1^{\pm}+u_2^{\mp 1}
\kop_1^{}\bos_2^{\pm}
\bos_3^{\mp}\right)\;,\\
\\
\ds \Rop_{123}^{}\; \bos_2^\pm\; \Rop_{123}^{-1} &=&  \ds
\bos_1^\pm \bos_3^\pm -
u_2^{\mp 1} \kop_1^{}\kop_3^{}\bos_2^{\pm}\;,\\
\\
\ds \Rop_{123}^{} \; \kop_2^{}\bos_3^{\pm}\; \Rop_{123}^{-1} &=&
\ds u_3^{\pm 1} \left( \kop_1^{}\bos_3^{\pm} + u_2^{\mp 1}
\kop_3^{}\bos_1^{\mp}\bos_2^{\pm}\right)\;,
\end{array}
\end{equation}
where
\begin{equation}
u_1=\frac{\lambda_3}{\lambda_2}\;,\quad u_2= -
\frac{1}{\lambda_1\mu_3}\;,\quad u_3=\frac{\mu_1}{\mu_2}\;,
\end{equation}
satisfies the adjoint tetrahedron equation in $\alg^{\otimes 6}$
and the quantum tetrahedron equation in proper
$\textrm{Rep}(\alg)^{\otimes 6}$ (spectral parameters
$\lambda_1,\mu_1,\dots \lambda_6,\mu_6$ for the tetrahedron
equation are free). Equations (\ref{Themap}) provide in addition
\begin{equation}
\Rop_{123}^{}\kop_1^{}\kop_2^{}\Rop_{123}^{-1}=\kop_1^{}\kop_2^{}\;,\quad
\Rop_{123}^{}\kop_2^{}\kop_3^{}\Rop_{123}^{-1}=\kop_2^{}\kop_3^{}\;.
\end{equation}
``Constant'' matrix $\rop_{123}$ corresponds to $u_1=u_2=u_3=1$.
In modular representation
\begin{equation}\label{modular}
q=\EXP^{\ii\pi\bb^2}\;,\quad \kop\;=\;-\ii
\EXP^{\pi\sigma\bb}\;,\quad \bb>0\quad\textrm{and}\quad
\sigma\in\mathbb{R}\;,
\end{equation}
the kernel of constant $\rop$-matrix is given by \cite{circular}
\begin{equation}\label{R-mod}
\begin{array}{l}
\ds \langle
\sigma_1^{}\sigma_2^{}\sigma_3^{}|\rop|\sigma_1'\sigma_2'\sigma_3'\rangle\;=\;
\delta_{\sigma_1^{}+\sigma_2^{},\sigma_1'+\sigma_2'}
\delta_{\sigma_2^{}+\sigma_3^{},\sigma_2'+\sigma_3'}
\sqrt{\frac{\varphi(\sigma_1)\varphi(\sigma_2)\varphi(\sigma_3)}
{\varphi(\sigma_1')\varphi(\sigma_2')\varphi(\sigma_3')}}
\\
\\
\ds
\EXP^{-\ii\pi(\sigma_1^{}\sigma_3^{}-\ii\eta(\sigma_1^{}+\sigma_3^{}-\sigma_2'))}\;
\int_{\mathbb{R}} \;du\; \EXP^{2\pi\ii u (\sigma_2'-\ii\eta)}\;
\frac{\varphi(u+\frac{\sigma_1'+\sigma_3'+\ii\eta}{2})\varphi(u+\frac{-\sigma_1-\sigma_3+\ii\eta}{2})}{\varphi(u+\frac{\sigma_1-\sigma_3-\ii\eta}{2})\varphi(u+\frac{\sigma_3-\sigma_1-\ii\eta}{2})}
\end{array}
\end{equation}
where $\varphi(z)$ is the ``non-compact quantum dilogarithm''
\cite{Faddeev:1995} defined by
\begin{equation}\label{dilog-noncompact}
\varphi(z)\;=\;\exp\left(\ds \frac{1}{4}\int_{\mathbb{R}+\ii 0}
\frac{\EXP^{-2\ii zw}}{\textrm{sinh}(w\bb)\textrm{sinh}(w/\bb)}\
\frac{dw}{w}\right)\;,\quad
\frac{\varphi(z-\ii\bb^{\pm1}/2)}{\varphi(z+\ii\bb^{\pm1}/2)}\;=\;1+\EXP^{2\pi
z\bb^{\pm1}}\;.
\end{equation}
Crossing parameter $\eta$ in (\ref{R-mod}) is given by
\begin{equation}\label{eta}
\eta=\frac{\bb+\bb^{-1}}{2}\;.
\end{equation}

In Fock space $(F^+)$ and anti-Fock space $(F^-)$ representations
representations
\begin{equation}\label{ff}
q=\EXP^{-\varepsilon}\;,\quad \kop\;=\; q^{n+1/2}\;,\quad
n=0,1,2,3\dots (F^+)\quad \textrm{or} \quad n=-1,-2,-3,\dots
(F^-)\;,
\end{equation}
the matrix elements of constant $\rop$-matrix are given by a
similar formula \cite{BS05,circular,supertetrahedron},
\begin{equation}\label{R-fock}
\begin{array}{l}
\ds \langle n_1^{}n_2^{}n_3^{}|\rop|n_1'n_2'n_3'\rangle\;=\;
\delta_{n_1^{}+n_2^{},n_1'+n_2'} \delta_{n_2^{}+n_3^{},n_2'+n_3'}
\prod_{i=1}^3 c_{n_i^{},n_i'}
\\
\\
\ds q^{n_1^{}n_3^{}+n_2'}\; \frac{1}{2\pi\ii} \oint
\;\frac{dz}{z^{n_2'+1}}\; \frac{(-q^{2+n_1'+n_3'}z;q^2)_\infty
(-q^{-n_1^{}-n_3^{}}z;q^2)_\infty}{(-q^{+n_1^{}-n_3^{}}z;q^2)_\infty(-q^{-n_1^{}+n_3^{}}z;q^2)_\infty}
\end{array}
\end{equation}
where
\begin{equation}
c_{n,n'}=\sqrt{\frac{(q^2;q^2)_{n'}}{(q^2;q^2)_n}}\quad
\textrm{if}\quad n=0,1,2,3\dots (F^+)
\end{equation}
or
\begin{equation}
c_{n,n'}=\sqrt{\frac{q^{n'(n'+1)}(q^2;q^2)_{-n-1}}{q^{n(n+1)}(q^2;q^2)_{-n'-1}}}
\quad \textrm{if}\quad n=-1,-2,-3,-4\dots (F^-)
\end{equation}
The \emph{clockwise} integration loop in (\ref{R-fock}) circles
all poles of the integrand but not includes $z=0$. Pochhammer's
symbols and Euler's quantum dilogarithm are defined by
\begin{equation}\label{dilog-compact}
(z;q^2)_n=(1-z)(1-q^2z)\cdots (1-q^{2(n-1)}z)\;,\quad
\frac{(-z/q;q^2)_\infty}{(-qz;q^2)_\infty}=1+z/q\;.
\end{equation}
Matrix (\ref{R-fock}) has the block-diagonal structure in
\begin{equation}
F_1^{\epsilon_1}\otimes F_2^{\epsilon_2}\otimes
F_3^{\epsilon_3}\;,\quad \epsilon_i=\pm\;,
\end{equation}
and thus it defines eight different $R$-matrices. Classical limits
of Fock and anti-Fock space representations (\ref{ff}) provide
\begin{equation}
0<k<1\quad \textrm{for $F^+$}\quad \textrm{and}\quad 1<k\quad
\textrm{for $F^-$}\;.
\end{equation}

Pre-factors and integrands in both (\ref{R-mod}) and
(\ref{R-fock}) have identical difference properties (leftmost
relations in (\ref{dilog-noncompact}) and (\ref{dilog-compact})),
the main difference is that there is a non-compact set of poles in
modular integrand and there is a compact set of poles in Fock
space integrand.

Advantage of the special case $u_1=u_2=u_3=1$ is that the constant
$\rop$ is the symmetric root of unity,
\begin{equation}
\rop_{123}^2=1\;,\quad \rop_{123}^\dagger = s_1s_2s_3
\rop_{123}^{} (s_1s_2s_3)^{-1}\;,
\end{equation}
where $s=1$ for modular representation and Fock representation
$F^+$ and
\begin{equation}
s=(-)^{n} \quad \textrm{for}\quad F^-\;.
\end{equation}
Factor $s$ takes into account the anti-unitarity
$(\bos^\pm)^\dagger=-\bos^{\mp}$ of anti-Fock representations. For
instance, matrix
\begin{equation}\label{faf}
\rop_{123}'\;=\;(-)^{n_2}\rop_{123}^{}\quad \textrm{in}\quad
F_1^{+}\otimes F_2^{-}\otimes F_3^{+}
\end{equation}
is Hermitian.

Spectral parameters in (\ref{Themap}) are given by ``external
field'' factors. All cases below correspond to spectral parameters
(\ref{u-reg1},\ref{u-reg2},\ref{u-reg3},\ref{u-reg4}) with
positive signs $\epsilon_1=\epsilon_2=\epsilon_3=1$.

\subsection{Regime 1} $R$-matrix of (\ref{Themap}) in modular
representation $\kop^2<0$ (\ref{modular}) and spectral parameters
of Regime 1 is given by
\begin{equation}\label{R-reg1}
\Rop_{123} =
\EXP^{-2\eta\phi_2\sigma_2}\rop_{123}\EXP^{2\eta\phi_1\sigma_1+2\eta\phi_3\sigma_3}\;.
\end{equation}
Kernel of (\ref{R-reg1}) is real and in the vicinity of
equilibrium point (\ref{k-tan}) it is positive with the asymptotic
\begin{equation}
\langle
\sigma_1\sigma_2\sigma_3|\Rop_{123}|\sigma_1\sigma_2\sigma_3\rangle
\sim \exp\left(-\frac{F_0}{\pi\bb^2}\right)\quad \textrm{as}\quad
\bb\to 0\;\;\textrm{and}\;\; -\ii\EXP^{\pi\bb\sigma_i}\to k_i,
\end{equation}
where free energy $F_0$ as function of $\phi_{1..3}$ is given by
(\ref{free-energy}). Presumably, partition function per site for
cubic lattice and $R$-matrix (\ref{R-reg1}) in physical regime
$0<\beta_j<\pi$ is
\begin{equation}
z\;=\;\exp\left(-\frac{4\eta^2}{\pi} F_0\right)
\end{equation}
for arbitrary $\eta>0$ (\ref{eta}).

\subsection{Regime 2} $R$-matrix of (\ref{Themap}) in modular
representation $\kop^2<0$ (\ref{modular}) and spectral parameters
of Regime 2 is given by
\begin{equation}\label{R-reg2}
\Rop_{123} =
\varrho^{-1}\EXP^{2\ii\eta\phi_2\sigma_2}\rop_{123}\EXP^{-2\ii\eta\phi_1\sigma_1-2\ii\eta\phi_3\sigma_3}\;,
\end{equation}
where $\varrho$ is a unitary constant multiplier. In the vicinity
$-\ii\EXP^{\pi\bb\sigma_i}\to k_i$ of ground state (\ref{k-tan})
for Regime 2 the kernel of constant $\rop$-matrix oscillates.
$R$-matrix (\ref{R-reg2}) is unitary and therefore it is the
building block for a Heisenberg evolution operator. However, a
spectral equation for the evolution operator is not yet known and
we can't rigorously deduce a relation between spectra of quantum
field theory and solitons and dispersion relation (\ref{ds-4}) of
classical field theory.

\subsection{Regime 3} A self-consistent quantum field theory for Fock space representations
corresponding to spectral parameters (\ref{u-reg3}) and
(\ref{k-reg3}) is defined by
\begin{equation}\label{R-reg3}
\Rop_{123} = \varrho^{-1}
\EXP^{\ii(\pi-\phi_2)n_2}\rop_{123}\EXP^{\ii\phi_1n_1+\ii\phi_3n_3}\;\;\textrm{in}\quad
F_1^{-}\otimes F_2^{+}\otimes F_3^{-}\;.
\end{equation}
Constant $\rop$-matrix in $F_1^{-}\otimes F_2^{+}\otimes F_3^{-}$
oscillates, operator (\ref{R-reg3}) is the unitary one for unitary
constant multiplier $\varrho$. A spectral equation for Heisenberg
evolution operator is not known either except for a special $1+1$
dimensional case and small occupation numbers \cite{Laser}. In the
same way as for Regime 2 we can not deduce rigorously relations
between quantum spectra and classical dispersion relation. Note
however an interesting feature of Regime 3: a self-consistency
prescribes a correspondence between signatures $\epsilon_i$ of
spectral parameters and a choice of representation $F^+$ with
$0<k<1$ or $F^-$ with $k>1$. Presumably, this correspondence
provides a proper physical interpretation of spectra of evolution
operators (see \cite{Laser} for $1+1$ dimensional case). A choice
of constant $\varrho$ both in Regimes 2 and 3 is also a subject of
proper physical interpretation.

\subsection{Regime 4} Curiously, the field-theoretical Regime 4
has no quantum field-theoretical counterpart; it corresponds to
divergent statistical mechanics.

$R$-matrix for Regime 4 (\ref{u-reg4}) is given by
\begin{equation}\label{R-reg4}
\Rop_{123}\;=\;\EXP^{\phi_2n_2}\rop_{123}'\EXP^{-\phi_1n_1-\phi_2n_2}\quad
\textrm{in}\quad F_1^{+}\otimes F_2^{-}\otimes F_3^{+}
\end{equation}
where $\rop'$ is defined by (\ref{faf}) and representation
$F_1^{+}\otimes F_2^{-}\otimes F_3^{+}$ is chosen in accordance
with (\ref{k-reg4}). Matrix elements of (\ref{R-reg4}) are
strictly positive and diverge as
\begin{equation}
\langle n|\Rop|n\rangle \sim q^{n_1n_2-n_1n_3+n_2n_3+\textrm{lower
terms}}\quad \textrm{as}\quad n_1,n_3\to \infty\;,\quad n_2\to
-\infty\;.
\end{equation}
A well-defined statistical mechanical lattice theory should
involve a compensation of quadratic exponential asymptotic. It is
possible via certain non-linear boundary conditions preserving the
integrability and involving extra three temperature-like
parameters.

In quasi-classical limit $q=\EXP^{-\varepsilon}\to 1$ diagonal
matrix element of $\rop'$ is given by
\begin{equation}\label{faf-ass}
\langle n|\rop'|n\rangle \;\sim\;
\EXP^{\phi_1n_1-\phi_2n_2+\phi_3n_3}\;
\exp\left(\frac{V_0}{\varepsilon}\right)\quad \textrm{as}\quad
\varepsilon\to 0\;,
\end{equation}
where finite $k_i=q^{n_i}$ define by (\ref{k-reg4}) and
(\ref{cos-reg4}) the hyperbolic triangle with dihedral angles
$\theta_i$ and positive sides $\phi_i$, $V_0$ is then given by
(\ref{V-reg4}). For so defined $\phi_i$ the field factor in
(\ref{R-reg4}) compensates the pre-exponent in (\ref{faf-ass}).
However, this ``classical equilibrium point''  has no relation to
a self-consistent quantum model.

\subsection{Regimes 2 and 3 as gauge field theories}
Well defined quantum field theories in Regimes 2 and 3 involve
Bose $q$-oscillators. However, the algebraic approach to the
quantum tetrahedron equations allows one to introduce Fermi
oscillators in addition to Bose ones \cite{supertetrahedron}.  All
fermionic $R$-matrices are even, they involve two fermions and one
boson. Both Fermionic and bosonic $R$-matrices are building blocks
of Heisenberg evolution operator. One can straightforwardly
consider an evolution of simple test states with small total
occupation numbers (in the Fock space representation for bosons).
The evolution produces a set of Feynmann diagrams on constant time
discrete surface (kagome lattice). In addition to simple
propagation, fermionic $R$-matrices are responsible for emissions
of bosons, decay of boson into fermion pair and annihilation of
fermion pair into a boson. Thus, the interpretation of quantum
field theories as gauge field theories, where the bosons are gauge
fields and fermions are matter field, is quite natural.
Presumably, a proper choice of spectral parameters provides also a
gap between bosonic ground state and fermionic ground state, thus
the spectral parameters are responsible in addition for fermionic
mass.

\section{Discussion: Algebraic curves of higher \emph{genera}}

Formulas (\ref{k-alg}) for the soliton solution  from the previous
sections formally coincide with those for a general complex
algebraic geometry (finite gap) solution:  $P_i,Q_i$ are divisors
on a genus $g$ algebraic curve $\Gamma_g$, $E$ is a prime form on
it, $\boldsymbol{f}$ is related to a point on
$\textrm{Jac}(\Gamma_g)$, $\Theta_{\nop}$ is a Theta-function:
\begin{equation}\label{Theta2}
\Theta_{\nop}=\Theta(I(\nop))\;,\quad I(\nop)=\boldsymbol{z}+
n_1\int_{Q_1}^{P_1}\boldsymbol{\omega}-n_2\int_{Q_2}^{P_2}\boldsymbol{\omega}+n_3\int_{Q_3}^{P_3}\boldsymbol{\omega}\;\in\;\textrm{Jac}(\Gamma_g),
\end{equation}
where $\boldsymbol{\omega}$ is a vector of Abel' holomorphic
differentials and $\boldsymbol{z}$ is an arbitrary point on the
Jacobian. Any of three-terms relation (\ref{themap}) is just the
Fay identity \cite{Mum83}. Expression for $I(\nop)$ in
(\ref{Theta2}) corresponds to a special case of homogeneous
divisors. Divisors $P_i,Q_i$ are not free, they are divisors of
three meromorphic functions,
\begin{equation}\label{meromorphic}
N\int_{Q_i}^{P_i}\boldsymbol{\omega}\;=\;0\quad\mod
(\pi,\pi\Omega)\;,\quad i=1,2,3,
\end{equation}
where $N$ is a size of cubic lattice,  and $\Omega$ is a period
matrix; equations (\ref{meromorphic}) provide the periodical
boundary conditions.

Soliton solution is not just a straightforward trigonometric limit
of algebraic geometry one since conditions (\ref{meromorphic}) are
relaxed, $P_i$ and $Q_i$ in general complex soliton solution are
free.

For the discrete time evolution system, the initial data of Cauchy
problem define uniquely the algebraic curve \cite{Korepanov:1995}
and thus selects the cases of soliton or finite gap dynamics.

For general finite gap dynamics the spectral parameters $u_i$ in
(\ref{themap}) are not uniquely defined. The reason is that
equations of motion (\ref{themap}) have gauge invariance. The
gauge transformation
\begin{equation}
a_{j,\nop}^{\pm}\;\to\;\xi_{j,\nop}^{\pm 1} a_{j,\nop}^{\pm}
\end{equation}
such that
\begin{equation}
\xi_{2,\nop+\e_2}\;=\;\xi_{1,\nop}\xi_{3,\nop}
\end{equation}
is equivalent to the transformation of spectral parameters
\begin{equation}
u_1\;\to\;u_1\frac{\xi_{1,\nop}}{\xi_{1,\nop+\e_1}}\;,\quad
u_2\;\to\;u_2\frac{\xi_{2,\nop+\e_2}}{\xi_{2,\nop}}\;,\quad
u_3\;\to\;u_3\frac{\xi_{3,\nop}}{\xi_{3,\nop+\e_3}}\;.
\end{equation}
Existence of homogeneous solution of equations of motion
(\ref{themap}) fixes the gauge group element and provides thus the
definition of $u_j$. Otherwise, parameters $u_j$ are irrelevant.

Based on the principles of quantum-classical correspondence, one
can conclude that the canonical quantization of $q$-oscillators
(\ref{qosc}) and the choice of the Hilbert space as the product of
local irreducible representations of $q$-oscillators corresponds
to the choice of soliton sector on classical equations of motion.
In particular, the condition of polynomial structure of
$Q$-operators for a nested Bethe Ansatz for Fock space
representations literally corresponds to factorization of spectral
curve. The finite gap sector must correspond thus to
\emph{another} quantization scheme -- a finite-gap quantization.

\bigskip

\noindent\textbf{Acknowledgements.} I am grateful to  V. Bazhanov,
R. Kashaev, V. Mangazeev and P. Vassiliou for valuable discussions
and fruitful collaboration.

%\bibliography{geometry}

\begin{thebibliography}{10}

\bibitem{circular}
V.~V. Bazhanov, V.~V. Mangazeev, and S.~M. Sergeev,
\emph{{Q}uantum geometry of
  3-dimensional lattices}, J. Stat. Mech. (2008), P07006, arXiv:0801.0129.

\bibitem{BS05}
V.~V. Bazhanov and S.~M. Sergeev, \emph{Zamolodchikov's
tetrahedron
  equation and hidden structure of quantum groups}, J. Phys. A \textbf{39}
  (2006), no.~13, 3295--3310.

\bibitem{BoSurBook}
A.~Bobenko and Yu. Suris, \emph{{D}iscrete differential geometry.
{C}onsistency
  as integrability.}, Monograph pre-published at
  http://www.arxiv.org/math/0504358, 2005.

\bibitem{BobenkoPinkall}
A. Bobenko and U. Pinkall, \emph{Discrete isothermic surfaces}, J.
  Reine Angew. Math. \textbf{475} (1996), 187--208.

\bibitem{Doliwa-circ}
A.~Doliwa, S.~V. Manakov, and P.~M. Santini,
  \emph{{$\overline\partial$}-reductions of the multidimensional quadrilateral
  lattice. {T}he multidimensional circular lattice}, Comm. Math. Phys.
  \textbf{196} (1998), no.~1, 1--18.

\bibitem{DoliwaSantini}
A. Doliwa and P.~M. Santini, \emph{Multidimensional quadrilateral
  lattices are integrable}, Phys. Lett. A \textbf{233} (1997), no.~4-6,
  365--372.

\bibitem{Faddeev:1995}
L.~D. Faddeev, \emph{Discrete heisenberg-weyl group and modular
group}, Lett.
  Math. Phys. \textbf{34} (1995), no.~3, 249--254.

\bibitem{KS98}
B.~G. Konopelchenko and W.~K. Schief, \emph{Three-dimensional
integrable
  lattices in {E}uclidean spaces: conjugacy and orthogonality}, R. Soc. Lond.
  Proc. Ser. A Math. Phys. Eng. Sci. \textbf{454} (1998), no.~1980, 3075--3104.

\bibitem{Korepanov:1994lomi4}
I.~G. Korepanov, \emph{A dynamical system connected with
inhomogeneous
  {$6$}-vertex model}, Zap. Nauchn. Sem. S.-Peterburg. Otdel. Mat. Inst.
  Steklov. (POMI) \textbf{215} (1994), no.~Differentsialnaya Geom. Gruppy Li i
  Mekh. 14, 178--196, 313.

\bibitem{Korepanov:1994lomi3}
\bysame, \emph{Hidden symmetries in a six-vertex model in
statistical physics},
  Zap. Nauchn. Sem. S.-Peterburg. Otdel. Mat. Inst. Steklov. (POMI)
  \textbf{215} (1994), no.~Differentsialnaya Geom. Gruppy Li i Mekh. 14,
  163--177, 312.

\bibitem{Korepanov:1994lomi2}
\bysame, \emph{Tetrahedron equation and the algebraic geometry},
Zap. Nauchn.
  Sem. S.-Peterburg. Otdel. Mat. Inst. Steklov. (POMI) \textbf{209} (1994),
  no.~Voprosy Kvant. Teor. Polya i Statist. Fiz. 12, 137--149, 262.

\bibitem{Korepanov:1994lomi1}
\bysame, \emph{Vacuum curves of {$\mathscr{L}$}-operators
associated with the
  six-vertex model}, Algebra i Analiz \textbf{6} (1994), no.~2, 176--194.

\bibitem{Korepanov:1995}
\bysame, \emph{Algebraic integrable dynamical systems, $2+1$
dimensional models
  on wholly discrete space-time, and inhomogeneous models on 2-dimensional
  statistical physics}, Adv. PhD Thesis, arXiv:solv-int/9506003, 1995.

\bibitem{Korepanov:1999tmp}
\bysame, \emph{Fundamental mathematical structures of integrable
models},
  Teoret. Mat. Fiz. \textbf{118} (1999), no.~3, 405--412.


\bibitem{KricheverNovikov}
I.~M. Kri{\v{c}}ever and S.~P. Novikov, \emph{Holomorphic vector
bundles over
  {R}iemann surfaces and the {K}adomcev-{P}etvia\v svili equation. {I}},
  Funktsional. Anal. i Prilozhen. \textbf{12} (1978), no.~4, 41--52.

\bibitem{Mum83}
D.~Mumford, \emph{Tata lectures on {T}heta. {I}, {I}{I}},
Birkh\"auser Boston
  Inc., Boston, Mass., 1983, 1984.

\bibitem{Q-Toda}
S. Pakuliak and S. Sergeev, \emph{Quantum relativistic {T}oda
chain
  at root of unity: isospectrality, modified {$Q$}-operator, and functional
  {B}ethe ansatz}, Int. J. Math. Math. Sci. \textbf{31} (2002), no.~9,
  513--553.

\bibitem{Sergeev:PN}
S.~Sergeev, \emph{{Q}uantum integrable models in discrete 2+1
dimensional
  space-time: auxiliary linear problem on a lattice, zero curvature
  representation, isospectral deformation of the
  {Z}amolodchikov-{B}azhanov-{B}axter model.}, Particles and Nuclei \textbf{35}
  (2004), 1051--1111.

\bibitem{Laser}
\bysame, \emph{Evolution operators for quantum chains}, J. Phys. A
\textbf{40}
  (2007), no.~8, F209--F213.

\bibitem{Melbourne}
S.~M. Sergeev, \emph{Quantization of three-wave equations}, J.
Phys. A
  \textbf{40} (2007), no.~42, 12709--12724.

\bibitem{supertetrahedron}
\bysame, \emph{{S}uper-tetrahdera and super-algebras},
arXiv:0805.4653, 2008.

\bibitem{SolitonBook}
V.~E. Zaharov, S.~V. Manakov, S.~P. Novikov, and L.~P.
Pitaevski{\u\i},
  \emph{Teoriya solitonov}, ``Nauka'', Moscow, 1980, Metod obratnoi zadachi.
  [The method of the inverse problem].

\bibitem{Zakharov:1973jetp}
V.~E. Zakharov and S.~V. Manakov, \emph{Resonant interaction of
wave packets in
  nonlinear media}, JETP Lett. \textbf{18} (1973), 243--245.

\end{thebibliography}
%\bibliographystyle{amsplain}

\def\cprime{$'$} \def\cprime{$'$}
\providecommand{\bysame}{\leavevmode\hbox
to3em{\hrulefill}\thinspace}
\providecommand{\MR}{\relax\ifhmode\unskip\space\fi MR }
\providecommand{\MRhref}[2]{%
  \href{http://www.ams.org/mathscinet-getitem?mr=#1}{#2}
} \providecommand{\href}[2]{#2}

\end{document}